\documentclass{aa}
\usepackage{txfonts}
\usepackage{graphicx}
\usepackage{natbib}

\usepackage{natbib}
\titlerunning{$\rm C_2$ in cool DQ white dwarfs}
\authorrunning{Kowalski}

\offprints{Piotr Kowalski, \email{kowalski@gfz-potsdam.de or pmkowalski@aol.com.
}}

\begin{document}

\title{The origin of peculiar molecular bands in cool DQ white dwarfs}
\author{Piotr M. Kowalski}

\institute{Helmholtz Centre Potsdam, GFZ - German Research Centre for Geosciences, Telegrafenberg, 14473 Potsdam, Germany}

\abstract{}
{
The DQ white dwarfs are stars whose atmosphere is enriched with carbon,
which for cool stars ($T_{\rm eff}<8000\rm \, K$) is indicated by the Swan bands of $\rm C_2$
in the optical part of their spectra. 
With decreasing effective temperature these molecular bands undergo a significant blueshift ($\sim 100-300 \AA$).
The origin of this phenomenon has been disputed over the last two decades and has remained unknown.
We attempt to address this problem by investigating the impact of dense helium on the spectroscopic properties of molecular carbon,
the electronic Swan band transition energy $T_{\rm e}$ and the vibrational frequency $\omega_{\rm e}$,
under the physical conditions encountered inside helium-rich, fluid-like atmospheres of cool DQ white dwarfs.
}
{
In our investigation we use a density functional theory based quantum mechanical approach.
}
{
The electronic transition energy $T_e$ increases monotonically with the helium density ($\Delta T_{\rm e}\rm\, (eV)\sim1.6 \, \it \rho \rm \, (g/cm^3)$).
This causes the Swan absorption to occur at shorter wavelengths compared with unperturbed $\rm C_2$.
On the other hand the pressure-induced increase in the vibrational frequency is insufficient to account for the observed Swan bands shifts.
Our findings are in line with the shape of the distorted molecular bands observed in DQp stars, but the predicted photospheric density required to
reproduce these spectral features is one order of magnitude lower than the one predicted by the current models. 
This indicates pollution by hydrogen or reflects incomplete knowledge of the properties of fluid-like atmospheres of these stars.
}
{
Our work shows that at the physical conditions encountered in the fluid-like atmospheres of cool DQ white dwarfs 
the strong interactions between $\rm C_2$ and helium atoms cause an increase in $T_e$, which should produce 
a blueward shift of the Swan bands. This is consistent 
with the observations and indicates that the observed Swan-like molecular bands are most likely the pressure-shifted bands of $\rm C_2$ 
}


\keywords{atomic processes -- dense matter -- stars: atmospheres -- stars: white dwarfs}

\maketitle

\section{Introduction}

\begin{figure}
\resizebox{\hsize}{!}{\rotatebox{270}{\includegraphics{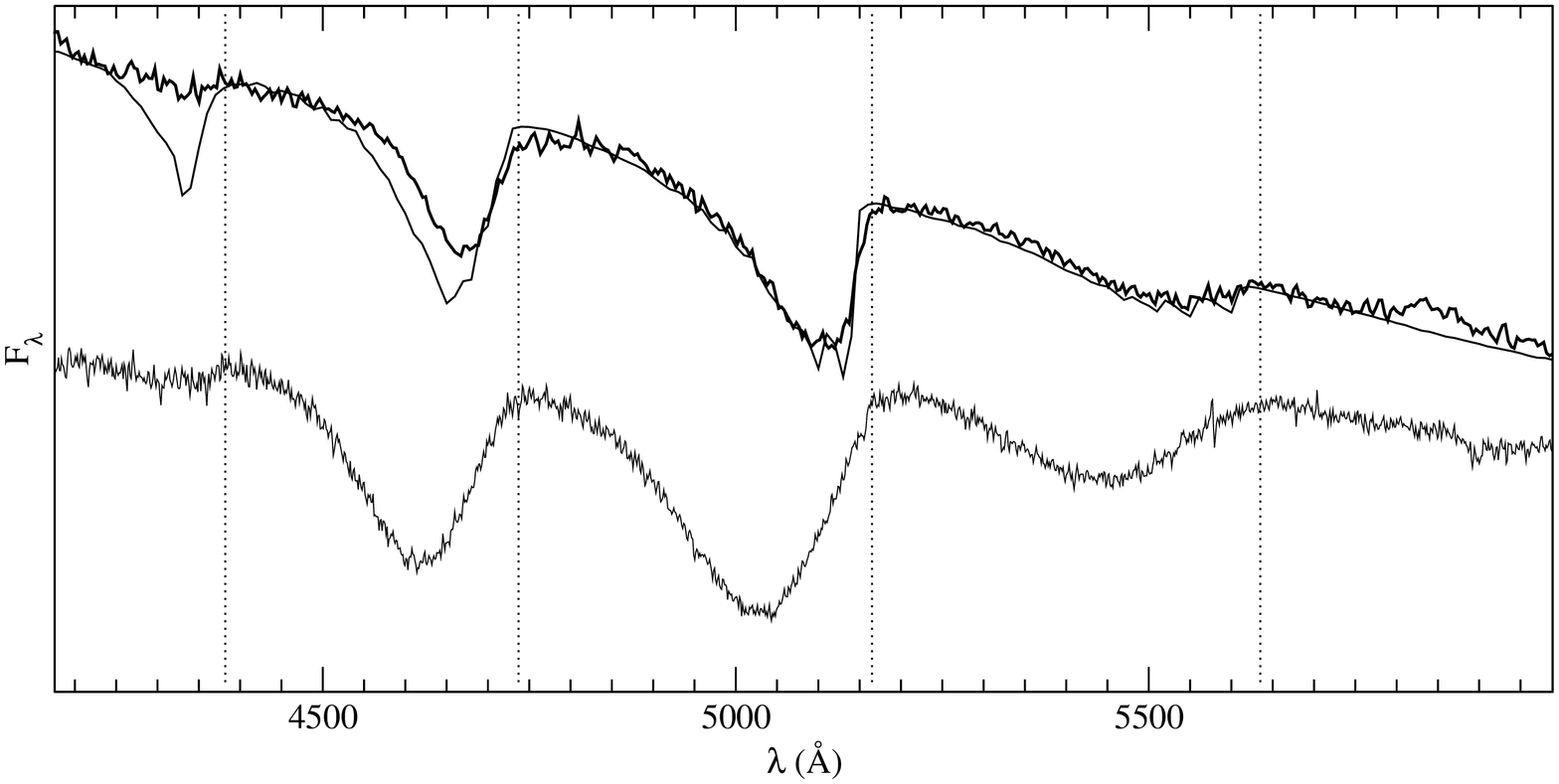}}}
\caption{Optical spectra of the white dwarfs LHS179 (DQ, upper panel) and LHS290 (DQp, lower panel) \citep{Bergeron97}.
The solid line represents the fit to the spectrum of LHS179. The parameters of the fits are: $T_{\rm eff}\rm=6500 \, K$, log g=8 (cgs),
and $\log \, \rm C/He=-6.4$. These values are in line with the atmospheric parameters derived in \citet{DB05}. The vertical dotted lines 
indicate the positions of the Swan bandheads for $\Delta\nu=\,+2,\,+1,\,0,\,-1$ from left to right respectively. The synthetic spectrum matches the entire known spectral energy distribution 
of the star (given by BVRIJHK photometry of \citet{Bergeron97}). \label{F1}}
\end{figure}

A DQ white dwarf is a star with a helium-dominated atmosphere enriched with carbon, which is believed to be dredged-up from the core by the deep 
helium convection zone \citep{P86}.  In the cool DQ white dwarfs ($T_{\rm eff}\rm <8000 \, K$) carbon shows its presence by the
$\rm C_2$ Swan bands in the optical spectrum. The two until now unexplained phenomena are observed in the local samples of these stars.
There is a cutoff in their cooling sequence at $T_{\rm eff}\rm \sim 6000 \, K$ \citep{DB05,KK06}
and an appearance of "peculiar" DQ stars  (following \citet{HM08} we call them DQp white dwarfs) at lower temperatures.
These peculiar stars show Swan-like bands that are blueshifted by 100 to 300 $\AA$ \citep{SBF95,Bergeron97,HM08}.
The representative spectra of normal and peculiar cool DQ stars are given in Fig. \ref{F1}.
In the detailed analysis of DQp stars \citet{SBF95,S99} and \citet{HM08} doubt the possibility that the shifts of the Swan bands 
observed in spectra of these stars are solely due to a strong magnetic field, which is not observed to be stronger than in ``normal" DQ stars. 
Observing identical shifts of the bands in all at that time known DQp stars, \citet{SBF95} concluded that a different molecular species could produce the observed bands.
Analyzing the abundances of chemical species in the H/He/C mixture under the physical conditions found in the atmospheres of helium-rich stars, 
they proposed the $\rm C_2H$ molecule as a source of the optical spectral features in the DQp stars.
The idea that hydrogen is responsible for the spectra of DQp stars is consistent with two findings regarding
the chemical evolution of cool white dwarf atmospheres.  
The already mentioned observation of DQ$\rightarrow$DQp transition indicates some sort of transformation of the physical or chemical properties of the atmospheres of these stars. 
Also the analyses 
of the local populations of white dwarfs 
by improved atmosphere models, which account for physics and chemistry of dense media, 
show the decrease in number of the helium-rich atmosphere stars with $T_{\rm eff}\rm <6000 \, K$ \citep{Bergeron97,Bergeron01,KS06}
and the appearance of white dwarfs with 
atmospheres highly enriched in hydrogen \citep{KS06,K08,K09a,K09b}.
Subsequent work on the coolest stars indicates that the stars at the end of the white dwarf cooling sequence posses pure hydrogen atmospheres \citep{H08,K10}.
Both these findings indicate that the accretion of hydrogen from the interstellar medium may change the composition of a helium-dominated atmosphere into a hydrogen-rich one,
and shows that the appearance of DQp stars may be somehow connected to this process. However, in the recent work of \citet{HM08} the authors 
ruled out the possibility that $\rm C_2H$ molecule or other carbon and hydrogen bearing species are responsible for the 
absorption in the optical spectra of DQp stars and concluded that $\rm C_2$ itself is a reasonable candidate for producing the observed features. This claim is supported by matching 
of the long-wavelength edges of the observed bands in DQ and DQp stars, as indicated in Fig. \ref{F1}. \citet{HM08} postulate that the Swan bands distortion could be produced by the high pressure or 
the high excitation of the rotational states of $\rm C_2$. This claim is however purely speculative and is not based on strong theoretical or experimental evidence.

\begin{figure}
\resizebox{\hsize}{!}{\rotatebox{270}{\includegraphics{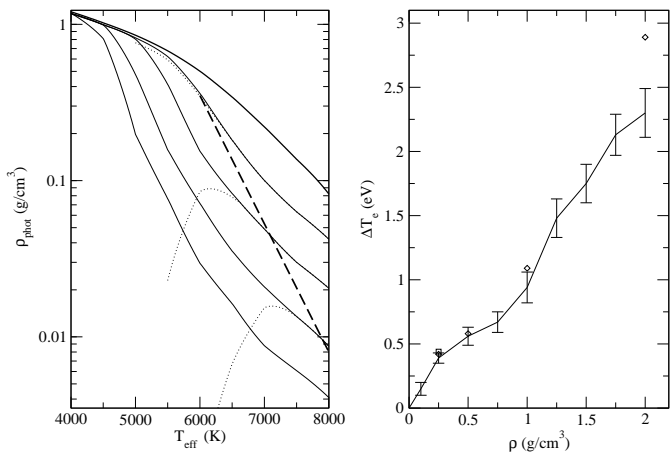}}}
\caption{Left panel: Density at the photosphere (the Rosseland mean optical depth $\tau_r=2/3$, solid lines)
 and at the optical depth of the $\Delta\nu=0$ Swan band formation region 
($\tau_{\lambda=5100\AA}=2/3$, dotted lines) in DQ white dwarfs of various He/C atmospheric compositions: 
$\log \rm He/C=-4,\,-5,\,-6,\,-7$ and pure-He from bottom to top.
The dashed line represents the photospheric density of the  
DQ star cooling sequence, assuming the carbon abundance derived by \citet{DB05}. No hydrogen bearing species are present in the models.
Right panel: The change in the electronic transition energy $\Delta T_e$ for the Swan bands transition in $\rm C_2$ as a function of helium density.
Error bars represent the results of calculations fixing inter-atomic distance in $\rm C_2$ at values derived for isolated $\rm C_2$. The square at $\rho=\rm 0.25 \, g/cm^3$ 
shows the result of the same calculation, but performed on the distributions of He atoms obtained from a quantum molecular dynamics simulation.
The diamonds represent the results obtained relaxing the inter-atomic distance in $\rm C_2$ to the equilibrium position. The temperature is $T=\rm 5802 \, K=0.5\,eV$.
\label{F2}}
\end{figure}

The majority of DQ star atmospheres are highly depleted with hydrogen, and the dominant constituent species is helium.
With cooling, an atmosphere of a helium white dwarf becomes more charge-neutral, less opaque and more dense at the photosphere.
Also the amount of carbon, which significantly contributes to the opacity by delivering free electrons, diminishes with the effective temperature (\citet{DB05,KK06}),
causing an increase in the photospheric density. The photospheric densities as a function of effective temperature for DQ white dwarfs with representative amounts of carbon are given in Fig. \ref{F2}.
The atmospheres of these cool stars ($T_{\rm eff}<8000 \rm \, K$) posses extreme densities (as high as few $\rm g/cm^3$) and represent a dense fluid, in which strong inter-particle 
interactions affect the chemistry and physics \citep{BSW95,KS04,Kowalski06,KSM07}.
Plotting the photospheric densities of a DQ star cooling sequence in Fig. \ref{F2}, which is based on the carbon abundances derived by \citet{DB05}, we see that the atmospheres of DQ stars reach fluid-like densities
and therefore high pressures at the temperatures of the observed DQ$\rightarrow$DQp transition. This suggests the pressure effects as a good candidate to be responsible for the spectral distortions observed in DQp stars.

In order to understand the origin of the spectral features observed in DQp stars, we investigate the impact of dense helium fluid on the spectroscopically 
important properties of $\rm C_2$ molecule. We performed {\it ab initio} calculations of $\rm C_2$ in dense helium, looking for changes in quantities that shape the optical spectrum
of $\rm C_2$, the electronic transition energy $T_{\rm e}$, and the vibrational frequency $\omega_{\rm e}$.

\section{Computational approach}

The most widely used method for the quantum mechanical treatment of dense, extended many-particle systems is the density functional theory (DFT, \citet{KH00}).
Over the last decade DFT methods 
proved to be very useful
in calculations of various chemical and physical properties of atoms, molecules, solid states and matter under high compression.
In our work we used one of the most common DFT version, the generalized gradient approximation (GGA) with PBE exchange-correlation functional \citep{Perdew2}.
The calculations were performed by the plane-wave DFT ABINIT code \citep{AB1} with the ultra-soft pseudopotentials \citep{DV90} (PAW method and the energy cutoff of $\rm 15 \, a.u.$).
In order to obtain the distributions of helium atoms around $\rm C_2$ we derived the $\rm He-C_2$ interaction potential 
using a combined DFT, molecular dynamics, and classical theory of fluids approach.
First we performed Car-Parrinello simulations \citep{CPMD2} of $\rm C_2$ emerged in dense helium at density of $0.5 \rm \, g/cm^{3}$
and the temperature of $\rm 5802 \rm \, K$ (equiv. of $0.5\rm\,eV$). The periodically repeated simulation box contained one $\rm C_2$ molecule and 64 He atoms.
As the ABINIT code is not well suited for these calculations, they were performed 
with the CPMD code \citep{CPMD2,CPMD3}, keeping the same psuedopotentials, DFT functional and energy cutoff.
We checked that using the same computational setups both DFT codes give the same answers for the total energies (within $\rm 10^{-5} a.u.$).
A trajectory of one million configurations was generated ($100 \rm \, ps$ long). The $\rm C_2-He$ interaction potential was derived by solving the Ornstein-Zernike equation 
in the Percus-Yevick approximation \citep{M}, taking as input the pair-distribution function obtained from the simulation. 
In order to calculate the energies of $\rm C_2$ in dense helium in $^3\Pi_u$ and $^3\Pi_g$ states, which constitute the Swan band transition, 
for a given $T$ and $\rho$ we calculated the ground state energies of a set of $50$ ionic configurations
obtained from the classical Monte Carlo simulations of helium fluid with one $\rm C_2$ molecule, performed using \citet{AT87}
Monte Carlo code with the derived $\rm C_2-He$ interaction potential and $\rm He-He$ interaction of \citet{RY}. We used this approach because 
an explicit quantum molecular dynamics (QMD) simulations would be computationally much more demanding,
and be undoable for extended systems with hundreds of atoms.
The relevant $^3\Pi_u$ and $^3\Pi_g$ states are constructed according to the electronic configurations given by \citet{Ballik}.
The symmetries of the five lowest KS orbitals of $\rm C_2$ molecule are: $\sigma_g(s)$,$\sigma_u(s)$,$\pi_u(p)$,$\pi_u(p)$,$\sigma_g(p)$, in agreement with \citet{Ballik}.
The lower $^3\Pi_u$ state ($(\sigma_g2s)^2(\sigma_u2s)^2(\pi_u2p)^3(\sigma_g2p)$) is obtained by fixing the occupation of spin-up orbitals as 11100 and spin-down 11111.
The upper $^3\Pi_g$ state ($(\sigma_g2s)^2(\sigma_u2s)(\pi_u2p)^3(\sigma_g2p)^2$) is obtained by fixing the occupation of spin-up orbitals as 10101 and spin-down 11111.
Then the Swan Band electronic transition energy is given as the difference of these two calculations. The obtained equilibrium separations for these two states, 
$2.49 \rm \, a.u.$, $2.37 \rm \, a.u.$ and the electronic transition energy $T_{\rm e}\rm=2.492 \, eV$ agree well with the experimental values 
($2.48 \rm \, a.u.$, $2.39 \rm \, a.u.$ and $2.39 \rm \, eV$ respectively, \citet{Ballik}), which justifies the usage of DFT.

The impact of dense helium on the vibrational frequencies of $\rm C_2$ was investigated by computing the vibrational density of states by the Fourier transform of the velocity
autocorrelation function \citep{EG94} derived from the velocities obtained from Car-Parrinello molecular dynamic runs for selected densities and $\rm C_2$ in a ground $\rm ^1X$ state.

In order to compute the atmosphere models we used our own stellar atmosphere code that accounts for various dense medium effects such as refraction, 
the non-ideal equation of state, and chemical equilibrium. The internal partition functions of carbon molecule is that of \citet{I81}.
For the carbon atom and ion we set the $T\rm=0\,K$ values. This approximation is justified because the electronic excitation 
energies of both carbon species are high, and for temperatures of a few thousand degrees the populations of the electronic excited levels are extremely small.
Hydrocarbon species are not considered.

\section{Results and discussion}  
It is well known that in helium fluid the interaction energies at the considered densities can be significant and may reach values of 
$\rm \sim eV´s$ (for instance the internal energy of atoms in dense helium is ~0.5 eV/atom at $1\rm \, g/cm^3$ and $T\rm=5802 \, K$). The interaction energy should in principle be higher for the species
in the higher excited states, because their electronic charge is located farther away from the center of the atom/molecule. This should lead to an increase 
in an electronic transition energy. We computed the electronic transition energy of the Swan transition in $\rm C_2$ that emerged in dense He at different densities and representative for the atmospheres of DQp stars
temperature $T=5802\rm\,K=0.5\,eV$.
The obtained shift in the electronic transition energy, $\Delta T_{\rm e}$ is given in Fig. \ref{F2}. With the increase in the density of helium, $T_{\rm e}$ increases and 
the increase for densities up to $0.25 \,\rm g/cm^3$ is linear and given by $\Delta T_{\rm e}\rm\, (eV)=1.6\,\rho_{\rm He} \, (g/cm^3)$.
This increase shows the smaller slope at densities between 0.5 and $1 \rm \, g/cm^3$ and a similar increase at higher densities. 
The increase in the electronic transition energy should produce 
a blueshift in the Swan band spectrum, which is qualitatively consistent with the shifts observed for DQp stars.
Positions of the band minima in the spectra of DQp stars imply a shift of $\rm \sim 0.08 \, eV$. 
This indicates that the bands form at a density of $\rm \sim 0.05 \, g/cm^3$, which is one order of magnitude smaller 
than the predicted photospheric densities in the models of these stars ($\rm \sim 0.4 \, g/cm^3$, Fig. \ref{F2}).

\begin{figure}
\resizebox{\hsize}{!}{\rotatebox{270}{\includegraphics{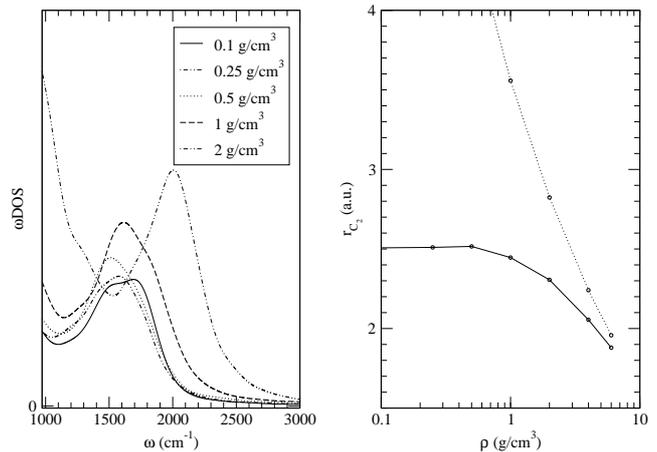}}}
\caption{Left panel: The vibrational density of states of $\rm C_2$ at different He densities. 
Right panel: the inter-particle distance in $\rm C_2$ as a function of helium density (solid line) obtained 
for $^1X$ state and the average inter-particle separation in dense helium (dotted line).
\label{F3}}
\end{figure}

On the other hand the change in the vibrational frequency is prominent, but appears at significantly higher densities.
The result in terms of the vibrational density of states of $\rm C_2$ in $\rm ^1X$ state is given in Fig. \ref{F3}.
The vibrational frequency is barely affected up to a density of $\rm 0.5 \, g/cm^{3}$ and shifts toward higher frequencies
at higher densities. This can be explained looking at the impact of compression on the mean inter-atomic separation in $\rm C_2$. In Fig. \ref{F3} we present 
the variation of this value as a function of density together with the mean inter-particle separation in the fluid helium.
The distance between carbon atoms that constitute the molecule remains unaffected up to a density of $\rm 0.5 \, g/cm^3$, and decreases slowly
at higher densities. The decrease is correlated with the mean inter-particle distance in fluid helium. The weak density-dependence of the inter-particle distance in $\rm C_2$ (for considered densities) 
justifies the usage of fixed inter-particle distance in the $\rm C_2$ molecule in the calculations of $T_{\rm e}$.
We also notice that to reproduce the observed shifts by the increase in the vibrational frequency, the difference in this increase should be at least $1000 \rm \, cm^{-1}$ ($\Delta E=0.5\hbar(\Delta \omega_{\rm e, upper}
-\Delta \omega_{\rm e, lower})$), which could occur at densities higher than $\rm 2 \, g/cm^3$. Potential rotational excitation, as proposed by \citet{HM08} for the reason of the shifts, 
is difficult to explain, because this is a temperature effect and should be stronger at higher temperatures, which is opposite to the temperature trend of the DQ$\rightarrow$DQp transition.
Therefore, we conclude that the dynamical motion of carbon atoms is unlikely to be responsible for the distortion of Swan bands. 
Our results show that the distortion of the $\rm C_2$ bands in DQp stars is caused by the pressure-induced increase in $T_{\rm e}$.

\begin{figure}
\resizebox{\hsize}{!}{\rotatebox{270}{\includegraphics{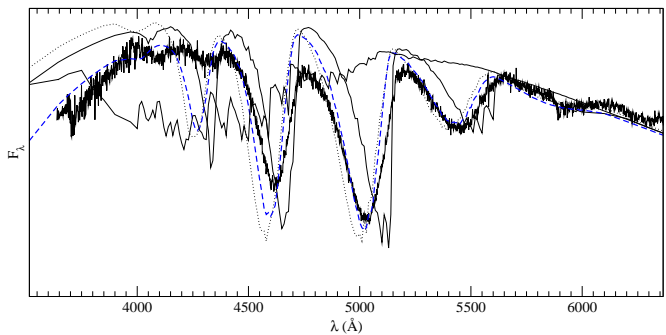}}}
\caption{Fits to the optical spectrum of the DQp white dwarf LHS290 (thick solid line).
All the fits are with $T_{\rm eff}\rm=5800 \, K$ and log g=8 (cgs).
The parameters of the fits are: H/He=0, $\rm C/He=1.25\cdot 10^{-7}$, $\Delta T_{\rm e}=0 \rm \, eV$ and $\Delta T_{\rm e}=1.6\, \rho_{\rm He} \rm \, eV$ (solid lines, $M_1$ and $M_2$); H/He=0, $\rm C/He=1.25\cdot 10^{-7}$, 
$\Delta T_{\rm e}=0.2\, \rho_{\rm He} \rm \, eV$ (dotted line, $M_3$); mixed composition $\rm He/H$ model, $\Delta T_{\rm e}=1.6\, \rho_{\rm He} \rm \, eV$, $\rm C/He=10^{-6}$ and $\rm H/He=6.75 \cdot 10^{-3}$ (dashed line, $M_4$).
All synthetic spectra well match the entire known spectral energy distribution of the star, as shown in Fig. \ref{F5}.
\label{F4}}
\end{figure}

\begin{figure}
\resizebox{\hsize}{!}{\rotatebox{270}{\includegraphics{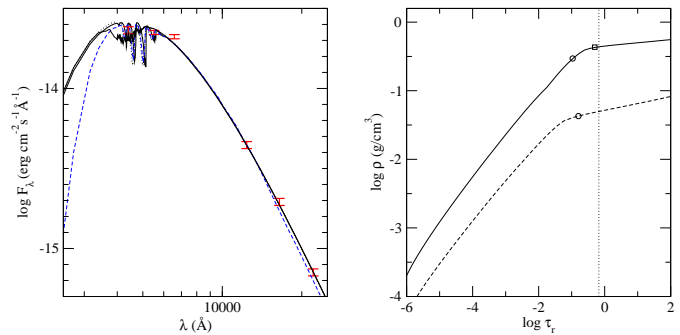}}}
\caption{Left panel: Fits to the entire spectral energy distribution of the DQp white dwarf LHS290 (thick solid line) 
given by BVIJHK photometry of \citet{Bergeron01} (bars). The parameters of the fits are the same as in Fig. \ref{F4};
Right panel: Density profiles of the discussed models. The vertical dotted line marks the photosphere ($\tau_r=2/3$). 
Circles indicate the optical depth at which $\Delta\nu=0$ band forms in models $M_1$, $M_3$ (upper curve) and $M_4$ (lower curve).
The square indicates the optical depth at which distorted bands form in model $M_2$.
\label{F5}}
\end{figure}

After computing the shifts in the electronic transition energy for the Swan band transition we attempted to reproduce the spectrum of a DQp star.
We did this by using the standard Swan band spectrum, shifted at a given atmospheric level by the derived value of $\Delta T_{\rm e}$. 
We note that although this is a quantitatively meaningful approximation, its validation would require the modeling of the absorption by $\rm C_2$ in dense helium, which is a complex task,
restricted by the limited applicability of quantum methods beyond DFT to many particle systems. In Fig. \ref{F4} we show the optical spectrum of the cool DQp white dwarf LHS290 \citep{Bergeron97} 
together with a set of synthetic spectra.
The overall spectral energy distribution of that star is best reproduced by models with $T_{\rm eff}\rm=5800 \, K$ (Fig. \ref{F5}), and we assume this temperature 
in our analysis and fix the gravity at $\log g=8 \,\rm (cgs)$. The C/He and H/He abundances are fitted to best reproduce the peaks of the $\Delta\nu=0,-1$ bands.
Assuming pure He atmosphere the strength of the Swan bands is reproduced with $\rm C/He=1.25\cdot 10^{-7}$, 
but with the computed correction for $T_{\rm e}$ the spectrum is far too distorted\footnote{We note that it resembles the spectrum observed for a star G240-72 in \cite{Bergeron97}, Fig. 30, which opens the possibility that this star ($T_{\rm eff}\rm=5640 \, K$, \citet{Bergeron97}) 
is not only magnetic, but also shows the extremely pressure-shifted Swan bands.}.
The observed spectrum can be fairly well reproduced by a model with photospheric density $\rm \sim 0.05 \, g/cm^3$, or assuming that the $T_{\rm e}$ dependence on the density is weaker, $\Delta T_{\rm e}\rm \sim0.2\rho_{\rm He}$.
The first case is realized by the addition of hydrogen to the atmosphere, which increases the opacity and lowers the photospheric density. The required amount of hydrogen is $\rm H/He=6.75 \cdot 10^{-3}$.
In both cases the observed spectrum is fairly well reproduced. The minima of the bands are blueshifted, as most of the absorption occurs close to the photosphere (Fig. \ref{F5}).
The long-wavelength parts of the bands resemble those of the standard Swan absorption, because part of the absorption occurs in the less dense upper atmospheric layers, where $\rm C_2$ is unperturbed.
We notice that the abundances of molecular carbon and the resulting strength of its molecular bands could also be
affected by high density, which could eventually impact the reported carbon abundances.

\section{Conclusions}

The so-called DQp stars represent a puzzle in the understanding of 
evolution of cool, helium-dominated white dwarf atmospheres. The DQ stars disappear at $T_{\rm eff}\rm \sim 6000 \, K$,
and few stars with apparently distorted Swan bands were detected at lower effective temperatures.
All explanation through the formation of different species, like $\rm C_2H$, magnetic fields, or roto-vibrational excitations failed to explain the
spectral features of these stars or definitely assign them as the distorted bands of $\rm C_2$. 
We show that 
the distortion of Swan bands originates in the pressure-induced increase in the electronic transition energy between states involved in the transition. 
This results in a blueshift of the molecular bands minima, and explains why the red edges of the bands match the spectra of normal DQ stars (Swan bands).
Our results, when applied to the current atmosphere models, predict Swan bands shifts that are too large compared with the observed ones. 
This indicates that the density at the photosphere of DQp stars does not excess $0.05 \rm \,g/cm^3$, and the input physics in the models or the understanding of 
the atmospheres of these stars, especially the pollution by hydrogen, requires further improvements.

\begin{acknowledgements}
I thank Sandy Leggett for providing me with the spectra of DQ and DQp stars, Didier Saumon for comments on the manuscript, the referee Patrick Dufour 
for constractive comments, suggestions and sharing the details of his DQ white dwarf atmosphere models,
and Richard Freedman for providing $\rm C_2$ opacities.
\end{acknowledgements}

\end{document}